\documentclass[sigconf,natbib=True, anonymous=False, review=False]{acmart}

\renewcommand\footnotetextcopyrightpermission[1]{} 
\AtBeginDocument{%
  \providecommand\BibTeX{{%
    \normalfont B\kern-0.5em{\scshape i\kern-0.25em b}\kern-0.8em\TeX}}}

\setcopyright{acmlicensed}
\copyrightyear{2024}
\acmYear{2024}
\acmDOI{XXXXXXX.XXXXXXX}

\acmConference[]{}{}
\acmSubmissionID{397}
\usepackage{orcidlink}
\usepackage{subcaption}
\usepackage{xcolor} 
\usepackage{multirow}
\usepackage[capitalise]{cleveref}

\everypar\expandafter{\the\everypar\loosness=-1}
\begin{document}


\title[A Reproducible Analysis of Sequential Recommender Systems]{A Reproducible Analysis of Sequential Recommender Systems}



\author{Filippo Betello}
    \email{betello@diag.uniroma1.it}
    \orcid{0009-0006-0945-9688}
    \affiliation{
        \institution{Sapienza University of Rome}
        \city{Rome}
        \country{Italy}}
    \authornote{These authors contributed equally to this research.} 

\author{Antonio Purificato}
    \email{purificato@diag.uniroma1.it}
    \orcid{0009-0006-0945-9688}
    \affiliation{
        \institution{Sapienza University of Rome}
        \city{Rome}
        \country{Italy}}
\authornotemark[1]

\author{Federico Siciliano}
    \email{siciliano@diag.uniroma1.it}
    \orcid{0000-0003-1339-6983}
    \affiliation{
        \institution{Sapienza University of Rome}
        \city{Rome}
        \country{Italy}}
    \authornotemark[1]

\author{Giovanni Trappolini}
    \email{trappolini@diag.uniroma1.it}
    \orcid{0000-0002-5515-634X}
    \affiliation{
        \institution{Sapienza University of Rome}
        \city{Rome}
        \country{Italy}}

\author{Andrea Bacciu}
    \email{bacciu@diag.uniroma1.it}
    \orcid{0009-0007-1322-343X}
    \affiliation{
        \institution{Sapienza University of Rome}
        \city{Rome}
        \country{Italy}}
    \authornote{Work done before joining Amazon.} 

\author{Nicola Tonellotto}
    \email{nicola.tonellotto@unipi.it}
    \orcid{0000-0002-7427-1001}
    \affiliation{
        \institution{University of Pisa}
        \city{Pisa}
        \country{Italy}}        

\author{Fabrizio Silvestri}
    \email{fsilvestri@diag.uniroma1.it}
    \orcid{0000-0001-7669-9055}
    \affiliation{
        \institution{Sapienza University of Rome}
        \city{Rome}
        \country{Italy}}



 
\renewcommand{\shortauthors}{Betello, Purificato, Siciliano et al.}

\begin{abstract}
    Sequential Recommender Systems (SRSs) have emerged as a highly efficient approach to recommendation systems. By leveraging sequential data, SRSs can identify temporal patterns in user behaviour, significantly improving recommendation accuracy and relevance.
   Ensuring the reproducibility of these models is paramount for advancing research and facilitating comparisons between them. Existing works exhibit shortcomings in reproducibility and replicability of results, leading to inconsistent statements across papers.
    \looseness -1 Our work fills these gaps by standardising data pre-processing and model implementations, providing a comprehensive code resource, including a framework for developing SRSs and establishing a foundation for consistent and reproducible experimentation.
    \looseness -1 We conduct extensive experiments on several benchmark datasets, comparing various SRSs implemented in our resource.
    We challenge prevailing performance benchmarks, offering new insights into the SR domain. For instance, SASRec does not consistently outperform GRU4Rec.
    On the contrary, when the number of model parameters becomes substantial, SASRec starts to clearly dominate all the other SRSs.
    This discrepancy underscores the significant impact that experimental configuration has on the outcomes and the importance of setting it up to ensure precise and comprehensive results.
    Failure to do so can lead to significantly flawed conclusions, highlighting the need for rigorous experimental design and analysis in SRS research.
    Our code is available at \url{https://github.com/antoniopurificato/recsys_repro_conf}.
\end{abstract}




\maketitle

\section{Introduction}

Recommender Systems (RSs) have been pivotal in enhancing user experience for many years by providing personalized recommendations across numerous online platforms \cite{7927889}.
Among these, Sequential Recommender Systems (SRSs) have received significant attention for their ability to consider the temporal dynamics inherent in user preferences to predict the next item a user will engage with \cite{quadrana2018sequence}.
Initially, SRSs used Markov chains to model user behaviour \cite{10.1145/1772690.1772773}, but faced challenges in capturing complex dependencies in extended user sequences \cite{Wang_2019}. The emergence of neural networks, especially recurrent structures \cite{cho2014learning,hidasi2016sessionbased}, marked a significant shift in this paradigm. Furthermore, the recent successes of transformer models \cite{vaswani2017attention} have inspired the development of novel SRS architectures \cite{sun2019bert4rec,kang2018selfattentive}.


Despite these advancements, the reproducibility\footnote{\label{footnote:reproducibility}We refer to the definition of \textit{reproducibility} of the ACM Artifact Review and Badging, version 1.1, available online at \url{https://www.acm.org/publications/policies/artifact-review-and-badging-current}.} of these models remains an open problem \cite{ferrari2019we,sun2020we}.
This issue is caused by several factors, mainly the lack of standardised benchmarks, uniform data pre-processing methods and consistent model implementations \cite{ferrari2020methodological, ferrari2021troubling, cremonesi2021progress}, which exacerbate the problem of reproducibility. Researchers frequently adopt different configurations, including various choices of hyperparameters and evaluation metrics \cite{hidasi2023effect, sun2020we, cremonesi2021progress}.
The variability of data processing methodologies across studies prevents direct comparability of results \cite{hidasi2023widespread}.
Differences in model implementations make it difficult to conduct a reliable and comparable evaluation of SRSs \cite{hidasi2023effect, petrov2022systematic}.
Finally, models are not consistently compared to the original settings described in the respective papers, introducing further ambiguity and preventing the establishment of a clear performance hierarchy \cite{petrov2022systematic}. 
The existence of different rankings of the proposed methods in each individual research paper hinders our ability to accurately assess real progress in the sequential recommendation domain.

We present EasyRec, a novel library dedicated to Sequential Recommender Systems (SRSs), designed to simplify data preprocessing and streamline model implementation. EasyRec provides a set of datasets and models, giving researchers and practitioners with an easy-to-use platform to facilitate experimentation and accelerate progress in the field of SRSs. Moreover, we conduct extensive experiments on multiple benchmark datasets, systematically comparing different SRSs implemented within our framework. This allows for a fair and unbiased evaluation, shedding light on the impact of crucial factors such as the length of the input sequence and the number of parameters.

The results of our study challenge existing performance benchmarks and provide new insights into the field of sequential recommendation. Surprisingly, our results show that SASRec does not consistently outperform GRU4Rec. However, an interesting shift occurs when the number of model parameters increases substantially: SASRec begins to clearly dominate all other SRSs.

Our contributions can be summarised as follows:
\begin{itemize}
    \item Provision of a comprehensive code resource, which allows different data pre-processing methods, facilitates the implementation of various SRSs, and ensures full experimental reproducibility.
    \item Extensive experiments on multiple datasets and models, reducing biases in performance comparisons.
    \item In-depth analysis of the results, providing new insights into different models and the impact of key hyperparameters;
    \begin{enumerate}
        \item Contrary to the prevailing notion in many articles, GRU4Rec consistently shows strong performance, often outperforming other SRSs.
        \item Performance gains for transformer-based models are more pronounced at higher embedding dimensions compared to GRU-based models, leading to SASRec outperforming all other models.
        \item The optimal input sequence length is related to the average user sequence length of each dataset, highlighting the importance of dataset-specific considerations.
        \item Transformer-based models exhibit exponential growth in size with increasing embedding dimensions, but performance improvements do not scale similarly. In addition, emissions are not only correlated with model size, but are significantly higher for models that require longer training times. 
    \end{enumerate}
\end{itemize}

\section{Related Work} \label{sec:related}

\subsection{Sequential Recommender Systems}

Markov chain-based techniques \cite{10.1145/1772690.1772773} are among the first approaches in SRSs, but they struggle to capture complex dependencies within long sequences \cite{Wang_2019}.

As a result, neural networks have recently emerged as the leading models for sequential recommendation. Recurrent Neural Networks \cite{10.1145/3159652.3159727} were among the first architectures introduced for this task. For instance, GRU4Rec \cite{hidasi2016sessionbased} is a recurrent neural network using Gated Recurrent Units (GRUs) \cite{cho2014learning}.
Another notable approach using GRUs is the Neural Attentive Recommendation Machine (NARM) \cite{narm}, which features a hybrid encoder-decoder architecture with a custom attention mechanism. This mechanism captures both local and global preferences in user-item interactions.

Advances in natural language processing have led to attention-based models. Self-Attentive Sequential Recommendation (SASRec) \cite{kang2018selfattentive} uses self-attention to discern the relevance of each item in the user's sequence. BERT4Rec \cite{sun2019bert4rec} instead exploits the Bidirectional Encoder Representations for Transformer (BERT) architecture \cite{devlin-etal-2019-bert}. Its bidirectional self-attention enables the model to obtain more accurate representations of user behaviour, improving recommendation performance. To avoid information leakage, BERT4Rec is trained on the Cloze task, i.e. predicting randomly masked items by jointly considering the left and right context in the input sequence.

A recent addition to the landscape of SRSs is CORE \cite{hou2022core}, which also integrates an attention mechanism. CORE offers two variants: one with custom attention computation and another using standard attention mechanisms. Both allow the model to weigh the contribution of each item in the input sequence, further refining the recommendation process.


\subsection{Reproducibility in SRSs}
Despite being a long-standing issue, reproducibility in SRSs has not received sufficient attention. \citet{ferrari2021troubling} sheds light on the problem, showing that only a fraction of top-quality work can be faithfully reproduced \cite{li2023repetition}. In support of this observation, \citet{hidasi2023effect} show that non-original implementations often produce inferior results than the original papers.
Even implementations found in common and widespread libraries, such as RecBole \cite{zhao2021recbole}, present this issue \cite{petrov2022systematic} and fail to faithfully reproduce reported results. Furthermore, even when results are reproducible, the lack of standardisation in faulty experimental setups (e.g. varying loss functions, user' sequence length, etc.) \cite{hidasi2023widespread} makes it impossible to establish a clear hierarchy between different models \cite{petrov2022systematic, klenitskiy2023turning, lin2019neural, ludewig2018evaluation} and to establish if we are really making progress in the SRS domain \cite{ferrari2019we}.
This inconsistency casts doubt on real progress in the SRS field \cite{ferrari2019we}.

Our work addresses these challenges by providing a comprehensive code resource using current versions of prominent Python libraries such as PyTorch and PyTorch Lightning. This framework ensures a faithful reproduction of the original paper results, demonstrating the robustness of our implementation. Furthermore, we conduct a rigorous comparative study under consistent conditions, allowing for a nuanced analysis of the hierarchy between SRSs. This approach contributes to a more reliable comparison process than current studies provide.

\section{Background}
The primary goal of sequential recommendation is to predict the next interaction within a sequence. Let's consider a collection of $n$ users, denoted as $\mathcal{U} \subset \mathbb{N^+}$ and a corresponding set of $n$ items denoted as $\mathcal{I} \subset \mathbb{N^+}$. Each user $u \in \mathcal{U}$ is characterised by a temporally ordered sequence $S_u = [s_1, \dots, s_{L_u}]$, where $s_i \in \mathcal{I}$ represents the $i$-th item the user has interacted with. Here, $L_u>1$ denotes the sequence length, which may vary from user to user.

A SRS, denoted as $\mathcal{M}$, takes as input the sequence up to the $L$-th item, expressed as $S_u^L = [s_1, \dots, s_L]$, and attempts to recommend the next item, $s_{L+1}$. The output of the recommendation, represented as $r^{L+1} = \mathcal{M}(S_u^L) \in \mathbb{R}^m$, is a score distribution over all items. This distribution is used to generate a ranked list of items representing the most likely interactions for user $u$ at step $L+1$. The following item $s_{L+1}$, is regarded as the \textit{positive} item, meaning that it is the item to be predicted. In contrast, all other items that should not be predicted at step $L+1$ are labeled as \textit{negative} items.

\paragraph{Input sequence length}
The choice of input sequence length $L$ is a crucial hyperparameter: opting for short sequences may provide insufficient information to the model, while excessively long sequences may inject irrelevant information. Given the varying number of interactions per user $L_u$, it is possible that $L_u$ is shorter than the chosen $L$. In such cases, the sequence is padded at the beginning with a dedicated identifier $pad \notin \mathcal{I}$, typically $0$.
Consequently, if $L_u < L$, the sequence is modified to $S_u^L = [pad, \dots, pad, s_1, \dots, s_{L_u}]$, with the number of padding instances equal to $L - L_u$. This approach ensures consistency in sequence lengths and allows for effective use of the model while accommodating variations in user interaction lengths.

\paragraph{Item Embedding}
The representation of items by identifiers alone lacks essential information about their inherent properties. To provide these representations with meaningful content, it is common practice to use a lookup table $\mathcal{E} : \mathcal{I} \rightarrow \mathbb{R}^d$. This table serves as a repository for embeddings, i.e. meaningful vectors that reside in a latent space with dimensionality $d$. This approach facilitates the transformation of item identifiers into rich, informative vectors that provide a nuanced and comprehensive representation of the main features of each item.

\section{Implementation Framework and Experimental Setup}

\subsection{Resource}
Our EasyRec library, is designed with several key features, listed below, that significantly enhance its utility for researchers and practitioners alike.

\paragraph{Reproducibility}
In our approach, we address two main challenges to improve the reproducibility of our experiments. First, we follow best practices for experimental seeding to ensure consistent and reliable results across multiple runs. Second, we specify the versions of the required Python packages, to prevent errors due to version mismatches, ensuring that our pipeline remains stable and predictable across different setups.
While the exact reproducibility of results is affected by non-negligible hardware and software factors, such as Python library versions and underlying hardware, we document all settings in our code repository\footnote{\label{footnote:code}\url{https://github.com/antoniopurificato/recsys_repro_conf}}.

\paragraph{Extensibility}
The resource is not restricted to the datasets, models or experimental setups used in this work. To include new datasets, it is sufficient to process them to adhere to the same format used in this work. For models, they can be easily added as PyTorch Module objects, with adjustments made in a configuration that specifies their input and output. The experimental setup is easily modifiable through a YAML file containing the entire configuration, enabling experimentation with scenarios not considered here. Further extensibility, described in detail in the code repository\footref{footnote:code}, includes the use of different loss functions, optimisers and more. 

\paragraph{Model implementations}
A common drawback of many existing implementations of popular SRSs is their overly complex and lengthy code, often using custom implementations of standardised methods. This makes it difficult to understand the model's operation and to assess its compliance with the mathematical principles underlying the methodology.
To address this issue, we undertook a comprehensive rewrite of these models, condensing them down to the most essential lines of code. Using PyTorch's pre-implemented methods, such as GRU and Attention, we were able to improve their clarity and comprehensibility.
Additionally, PyTorch implementations are widely adopted, standardized, and rigorously maintained, reducing the likelihood of errors and ensuring timely resolution of any issues.
By aligning with established PyTorch practices, our code stands to benefit from ongoing improvements in the PyTorch ecosystem. In addition, our library supports wandb\footnote{\url{https://wandb.ai/site/}}, which can be used to track experiments and perform hyperparameter optimization.

\paragraph{Running experiments}
The software environment can be reproduced by taking the Python library versions specified in our library requirements.
A YAML file contains the experiment configuration, specifying all hyperparameters.
A Python script reads this configuration and streamlines model training.
Detailed information can be found in the code repository\footref{footnote:code}.

\begin{table}[t!]
    \caption{Dataset statistics after pre-processing; users and items not having at least 5 interactions are removed. Avg. and Med. refer to the Average and Median of $\frac{\mathrm{Actions}}{\mathrm{User}}$, respectively.}
    \resizebox{\columnwidth}{!}{
      \begin{tabular}{l||ccc|ccc}
        \toprule
        Name & Users & Items & Interactions &  Density & Avg. & Med. \\ 
        \midrule
        Beauty & 1,274 & 1,076 & 7,113 & 0.519 & 5,58 & 5\\
        FS-NYC & 1,083 & 9,989 & 179,468 & 1.659 & 165 & 116\\
        FS-TKY & 2,293 & 15,177 & 494,807 & 1.421 & 215 & 146\\
        ML-100k & 943 & 1,349 & 99,287 & 7.805 & 105 & 64 \\
        ML-1M &  6,040 & 3,416 & 999,611 & 4.845 & 165 & 96\\
        ML-20M & 138,493 & 18,345 & 19,984,024 & 0.787 & 144 & 68\\
        \bottomrule
      \end{tabular}
    }
\label{tab:dataset_info}
\end{table}

\subsection{Datasets}
Our dataset collection includes the following:
\begin{itemize}
    \item Amazon\footnote{\url{https://jmcauley.ucsd.edu/data/amazon/}}: These datasets consist of product reviews collected from Amazon.com by \citet{10.1145/2766462.2767755}. The data are organized into distinct datasets based on Amazon's primary product categories.
    For our study, we focus on the ``Beauty'' category (Beauty).
    \item Foursquare\footnote{\url{https://sites.google.com/site/yangdingqi/home/foursquare-dataset}}: These datasets contain check-ins collected over a period of approximately ten months \cite{6844862}. We use the New York City (FS-NYC) and Tokyo (FS-TKY) versions.
    \item MovieLens\footnote{\url{https://grouplens.org/datasets/movielens}}: The MovieLens dataset \cite{10.1145/2827872} is widely recognized as a benchmark for evaluating recommendation algorithms. We utilize three versions: MovieLens 20M (ML-20M), MovieLens 1M (ML-1M) and MovieLens 100k (ML-100k).
\end{itemize}

The statistics of these datasets are presented in \cref{tab:dataset_info}.

\begin{figure*}[ht!]
    \includegraphics[width=0.9\textwidth]{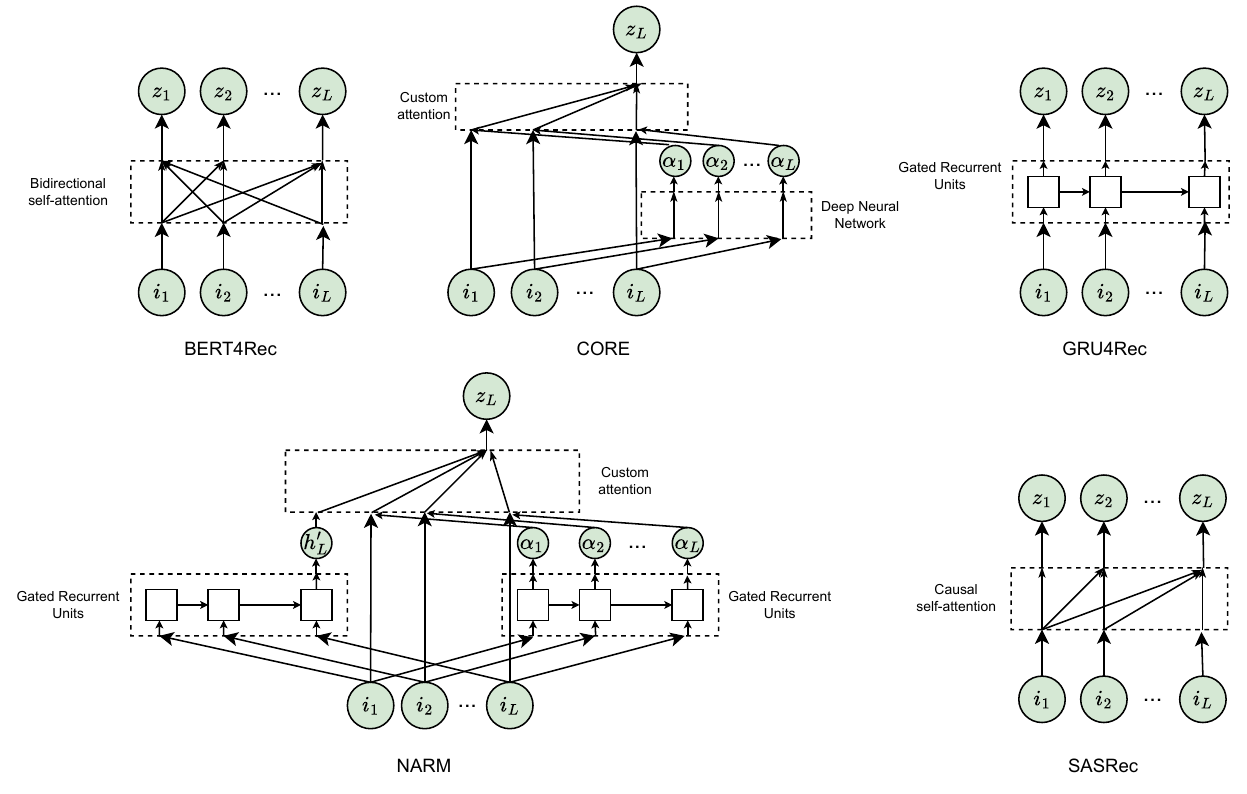}
    \caption{A visual depiction illustrating the core functionality of the considered SRSs. The input items ${i_1, i_2, \dots, i_L}$ are processed to generate the final representations ${z_1, z_2, \dots, z_L}$, which are utilized to generate predictions at steps $1, 2, \dots, L$, respectively. Intermediate representations ${h_1, h_2, \dots, h_L}$ are also present for some models.}
    \Description[A visual depiction illustrating the core functionality of the considered SRSs.]{For each SRS, the input items are processed to generate the final representations, which are utilized to generate predictions at steps $1, 2, \dots, L$, respectively.}
    \label{fig:models}
\end{figure*}

\paragraph{Data pre-processing}
Our pre-processing strategy follows established practices, such as treating ratings as implicit, i.e. use all interactions without considering the rating value, and removing users and items with fewer than 5 interactions \cite{kang2018selfattentive, sun2019bert4rec, petrov2022systematic}.
For testing, as in \cite{sun2019bert4rec, kang2018selfattentive}, we keep the last interaction for each user, while for the validation set, the second to last action is retained. All remaining interactions contribute to the training set.
Each observed user-item interaction is treated as a positive instance, and negative items are sampled from items not previously consumed by the user.

\subsection{Models} \label{sec:models}
We employ five sequential recommendation models in our experiments, each based on seminal papers widely cited in the literature. While a more detailed discussion of these models is provided in \cref{sec:related}, here we offer a succinct overview of their differences:

\begin{itemize}
    \item BERT4Rec \cite{sun2019bert4rec}: This model is based on the BERT architecture, enabling it to capture complex relationships in user behaviour sequences through bidirectional self-attention.
    \item CORE \cite{hou2022core}: it introduces an attention mechanism that enables the model to weigh the contribution of each item in the input sequence, enhancing recommendation accuracy.
    \item GRU4Rec \cite{hidasi2016sessionbased}: This model utilizes GRUs to capture temporal dependencies in user-item interactions.
    \item NARM \cite{narm}: it also leverages GRUs, but incorporates a custom attention mechanism to capture both local and global preferences in user-item sequences.
    \item SASRec \cite{kang2018selfattentive}: this model is characterized by its use of self-attention mechanisms, allowing it to discern the relevance of each item within the user's sequence.
\end{itemize}

It's crucial to distinguish between sequence-to-sequence models (GRU4Rec, SASRec, BERT4Rec) and sequence-to-item models (NARM, CORE). Sequence-to-sequence models analyze input sequences in detail, generating predictions at each step, addressing issues like vanishing gradients in long sequences \cite{pascanu2013difficulty}. They require fewer samples as each sequence generates multiple predictions, effectively increasing training capacity. Conversely, sequence-to-item models process the entire input sequence but only predict at the last step, reducing training capacity compared to sequence-to-sequence models. The original works \cite{narm, hou2022core} likely counter this by using shorter input lengths, reducing the need for a large number of training samples.
A schematic diagram illustrating the functioning of each model is provided in \cref{fig:models}.
\begin{table*}[!ht]
\caption{Comparison using MovieLens 1M dataset. RecBole$^a$ represents a model trained until it matches our metrics, while RecBole$^b$ refers to the model trained for the same time as ours.}
\begin{tabular}{cccccccccc}
\toprule
 \textbf{Model} & \multicolumn{3}{c}{\textbf{GRU4Rec}} & \multicolumn{3}{c}{\textbf{SASRec}} & \multicolumn{3}{c}{\textbf{BERT4Rec}} \\
 \cmidrule(lr){2-4} \cmidrule(lr){5-7} \cmidrule(lr){8-10} 
 \textbf{Implementation} & \textit{EasyRec} & \textit{RecBole$^a$} & \textit{RecBole$^b$} & \textit{EasyRec} & \textit{RecBole$^a$} & \textit{RecBole$^b$} & \textit{EasyRec} & \textit{RecBole$^a$} & \textit{RecBole$^b$} \\ \midrule
\textbf{NDCG@10} & 0.56 & 0.56 & 0.48  &  0.51 & 0.51  & 0.45 & 0.49 &  0.49 &  0.43  \\
\textbf{Recall@20} & 0.88 & 0.88 & 0.85 & 0.86 & 0.86 & 0.84 & 0.86 & 0.86 & 0.83 \\
\textbf{Time}  & 28min & 1330min  & 32min & 38min & 360min & 48min  & 51min & 220min & 68min  \\ \bottomrule
\end{tabular}
\label{tab:comparison_libraries}
\end{table*}
\paragraph{Training}
A common loss function for training SRSs \cite{kang2018selfattentive, 10.1145/3159652.3159656} is Binary Cross Entropy (BCE) \cite{good1952rational}. While some studies explore variations in training methods \cite{hidasi2016sessionbased, narm}, our primary focus is on evaluating the architecture itself, with the aim of isolating its capabilities from the effects of different loss functions.

We trained all models for $400$ epochs, and for each model, we selected the checkpoint from the epoch where it achieved the highest performance on the validation set.
We also utilize the same optimiser, batch size, number of negative items per training and testing instance, and other relevant parameters. By maintaining consistency in these aspects, we aim to eliminate potential confounding factors and facilitate a clear assessment of architectural differences.
For ease of reproducibility and transparency, all relevant parameters and configurations are carefully documented and made available in our code repository \footref{footnote:code}. All experiments were performed on a single NVIDIA RTX A6000 with 10752 CUDA cores and 48 GB of VRAM.

\subsection{Metrics}
To evaluate the performance of sequential recommendation algorithms, we use four widely used metric, also common in Information Retrieval (IR): Precision, Recall, NDCG and MAP.

\paragraph{Tracking Emissions}
In the Glasgow Agreement \cite{hunter2021glasgow} climate change highlights the international commitment to reduce CO\textsubscript{2} emissions. This is especially important in the field of Artificial Intelligence, where training GPUs has a significant environmental impact. The energy consumption associated with these computational processes contributes significantly to CO\textsubscript{2} emissions, exacerbating climate change \cite{patterson2021carbon}. It is therefore our responsibility to raise awareness of this issue.

In this study, we use CodeCarbon\footnote{\url{https://codecarbon.io}} \cite{codecarbon}, a tool designed to track the power consumption of both CPUs and GPUs. This allows us to measure carbon dioxide equivalent (CO\textsubscript{2}-eq), a widely accepted standard to monitor emissions of various greenhouse gases \cite{kim2009measurement}. CO\textsubscript{2}-eq facilitates the comparison of greenhouse gas emissions by converting quantities of different gases into an equivalent amount of CO\textsubscript{2}, based on their respective global warming potentials.

\subsection{Comparisons}
The importance of our resource lies primarily in the fact that it fills a notable gap in the existing literature and libraries. In particular, Elliot \cite{anelli2021elliot} and DaisyRec \cite{sun2020we} lack SRS models, while Cornac \cite{salah2020cornac} and RecPack \cite{recpack2022} provide only GRU4Rec among newer models. Although ReChorus \cite{wang2020rechorus} supports many models, it lacks one of the most cited ones (BERT4Rec) and modern ones. Conversely, RecBole \cite{zhao2021recbole} stands out as a comprehensive framework that is widely used in the field. However, it requires the creation of a subclass of its classes to develop new models or datasets. In contrast, our approach allows the extension of the PyTorch standard class for all new models. Furthermore, existing literature highlights suboptimal performance metrics, timing, and implementation quality in RecBole \cite{hidasi2023effect, petrov2022systematic}. Recbole's performance, evaluated on the same number of epochs, is compromised by performing many iterations per epoch due to their data augmentation method \footnote{\url{https://recbole.io/docs/recbole/recbole.data.dataset.sequential_dataset.html}}. Moreover, they did not implement BCE loss, which is used in several works \cite{kang2018selfattentive}, as also indicated on the official GitHub repository \footnote{\url{https://github.com/RUCAIBox/RecBole/issues/1667}}, so their implementations are not loyal to the originals. We changed their loss to be consistent with the original works. \cref{tab:comparison_libraries} shows the comparison between our implementation and that of RecBole using ML-1M and the three most common and widely used models in the literature: SASRec, GRU4Rec, and BERT4Rec, in terms of metrics and time. By RecBole$^a$, we refer to models that are tracked until they reach our metrics, while by RecBole$^b$ we refer to models that are tracked for our own time. The experimental setup is the same as described in \cref{sec:base-performance}. We outperform RecBole in every model.

\section{Results}

\begin{table*}[!ht]
\caption{Results of the proposed models in terms of Precision@K (P@K), Recall@K (R@K), NDCG@K and MAP@K, with \mbox{K $\mathbf{\in \{10,20\}}$}. \textbf{Bold} denotes the best model for a dataset by the metric in the main group, \underline{underlined} the second best. $^\dagger$ indicates a statistically significant result with $\textbf{p}$-value $\textbf{< 0.01}$.}
\begin{tabular}{ll|cccc|cccc}
\toprule
\textbf{Dataset} & \textbf{Model} & \textbf{P@10} & \textbf{R@10} & \textbf{NDCG@10} & \textbf{MAP@10} & \textbf{P@20} & \textbf{R@20} & \textbf{NDCG@20} & \textbf{MAP@20} \\
\midrule
\multirow{5}{*}{\textbf{Beauty}} & \textbf{BERT4Rec} & \textbf{0.0707} & \textbf{0.7072} & \textbf{0.6828} & \underline{0.1991} & 0.0362 & 0.7237 & \underline{0.6848} & \textbf{0.1233} \\
& \textbf{CORE} & 0.0695 & 0.6954 & 0.6594 & 0.1926 & \underline{0.0366} & \underline{0.7316} & 0.6673 & 0.1199 \\
 & \textbf{GRU4Rec} & \underline{0.0704} & \underline{0.7041} & \underline{0.6823} & \textbf{0.1992} & 0.0360 & 0.7198 & \textbf{0.6855} & \textbf{0.1233} \\
 & \textbf{NARM} & 0.0695 & 0.6954 & 0.6701 & 0.1957 & \textbf{0.037} & \textbf{0.7394} & 0.6754 & 0.1212 \\
& \textbf{SASRec} & 0.0703 & 0.7033 & 0.6650 & 0.1942 & 0.0364 & 0.7276 & 0.6703 & 0.1209 \\
 \midrule
\multirow{5}{*}{\textbf{FS-NYC}} & \textbf{BERT4Rec} & \textbf{0.0789} & \textbf{0.7895} & \textbf{0.6689} & \textbf{0.1909} & \textbf{0.0430} & \textbf{0.8596} & \textbf{0.6856} & \textbf{0.1230} \\
 & \textbf{CORE} & 0.0510 & 0.5097 & 0.4237 & 0.1206 & 0.0297 & 0.5937 & 0.4453 & 0.0785 \\
 & \textbf{GRU4Rec} & \underline{0.0752} & \underline{0.7516} & \underline{0.6409} & \underline{0.1837} & \underline{0.041} & \underline{0.8199} & \underline{0.6593} & \underline{0.1179} \\
  & \textbf{NARM} & 0.0338 & 0.338 & 0.2310 & 0.0629 & 0.0222 & 0.4432 & 0.2583 & 0.0445 \\
 & \textbf{SASRec}  & 0.0730 & 0.7304 & 0.6149 & 0.1760 & 0.0407 & 0.8135 & 0.6353 & 0.1137 \\
 \midrule
\multirow{5}{*}{\textbf{FS-TKY}}  & \textbf{BERT4Rec} & \underline{0.0816} & \underline{0.8155} & \underline{0.6721} & \underline{0.1921} & \textbf{0.0437$^\dagger$} & \textbf{0.8735$^\dagger$} & \underline{0.6861} & \underline{0.1242}\\
 & \textbf{CORE}& 0.0578 & 0.5778 & 0.4753 & 0.1351 & 0.0328 & 0.6568 & 0.4947 & 0.0882 \\
& \textbf{GRU4Rec} & 0.0772 & 0.7719 & 0.6259 & 0.1785 & 0.0422 & 0.8443 & 0.6448 & 0.1163 \\
 & \textbf{NARM} & 0.0443 & 0.4431 & 0.3121 & 0.0858 & 0.0274 & 0.5473 & 0.3377 & 0.0595\\
 & \textbf{SASRec}  & \textbf{0.0820$^\dagger$} & \textbf{0.8203$^\dagger$} & \textbf{0.6896$^\dagger$} & \textbf{0.1981$^\dagger$} & \underline{0.0435} & \underline{0.8692} & \textbf{0.7023$^\dagger$} & \textbf{0.1274$^\dagger$} \\
 \midrule
 \multirow{5}{*}{\textbf{ML-100k}}  & \textbf{BERT4Rec} & 0.0549 & 0.5493 & 0.3075 & 0.0785 & 0.0363 & 0.7253 & 0.3499 & 0.0604 \\
  & \textbf{CORE}& 0.0352 & 0.3521 & 0.1886 & 0.0478 & 0.025 & 0.5005 & 0.2250 & 0.0378 \\
   & \textbf{GRU4Rec} & \textbf{0.0643} & \textbf{0.6426} & \textbf{0.3736} & \textbf{0.0986} & \textbf{0.0393} & \textbf{0.7869} & \textbf{0.4138} & \textbf{0.073} \\
   & \textbf{NARM} & 0.0273 & 0.2725 & 0.1411 & 0.0352 & 0.022 & 0.4401 & 0.1813 & 0.0291 \\
 & \textbf{SASRec} & \underline{0.0602} & \underline{0.6023} & \underline{0.3354} & \underline{0.0861} & \underline{0.0383} & \underline{0.7667} & \underline{0.3754} & \underline{0.0659}\\
 \midrule
\multirow{5}{*}{\textbf{ML-1M}}  & \textbf{BERT4Rec} & 0.0752 & 0.7517 & 0.4944 & 0.1357 & 0.0428 & 0.8561 & 0.5214 & 0.0949\\
& \textbf{CORE}& 0.0457 & 0.4571 & 0.2601 & 0.0669 & 0.0312 & 0.6238 & 0.3029 & 0.0517 \\
 & \textbf{GRU4Rec}  & \textbf{0.0788$^\dagger$} & \textbf{0.7876$^\dagger$} & \textbf{0.5614$^\dagger$} & \textbf{0.1572$^\dagger$} & \textbf{0.0438$^\dagger$} & \textbf{0.8758$^\dagger$} & \textbf{0.5837$^\dagger$} & \textbf{0.1065$^\dagger$}\\
 & \textbf{NARM} & 0.0428 & 0.4285 & 0.2398 & 0.0611 & 0.0303 & 0.6051 & 0.2840 & 0.0480 \\
 & \textbf{SASRec}  & \underline{0.0761} & \underline{0.7614} & \underline{0.5063} & \underline{0.1396} & \underline{0.0431} & \underline{0.8614} & \underline{0.5319} & \underline{0.0971} \\
 \midrule
 \multirow{5}{*}{\textbf{ML-20M}}  & \textbf{BERT4Rec} & 0.0942 & 0.9419 & \underline{0.7381} & \underline{0.2118} & 0.0489 & 0.9773 & \underline{0.7471} & \underline{0.1381}\\
& \textbf{CORE} & 0.0786 & 0.7863 & 0.5269 & 0.1460 & 0.0451 & 0.9019 & 0.5578 & 0.1012\\
 & \textbf{GRU4Rec}  & \textbf{0.0952$^\dagger$} & \textbf{0.9521$^\dagger$} & \textbf{0.7741$^\dagger$} & \textbf{0.2233$^\dagger$} & \textbf{0.0491$^\dagger$} & \textbf{0.9812$^\dagger$} & \textbf{0.7817$^\dagger$} & \textbf{0.1440$^\dagger$} \\
 & \textbf{NARM}  & 0.0871 & 0.8706 & 0.6172 & 0.1736 & 0.0468 & 0.9355 & 0.6339 & 0.1172\\
 & \textbf{SASRec} & \underline{0.0943} & \underline{0.9432} & 0.7284 & 0.2086 & \underline{0.0490} & \underline{0.9807} & 0.7380 & 0.1366 \\
 \bottomrule
\end{tabular}
\label{tab:main_table_results}
\end{table*}

Our experiments aim to address the following research questions:

\textbf{RQ1}: What is the hierarchy of current SRSs and is it consistent with the existing literature?

\textbf{RQ2}: How does the performance of SRSs vary with changes in (a) embedding size and (b) input sequence length?

\textbf{RQ3}: Using a more appropriate comparison based on the total number of parameters, what insights can be gained about the hierarchy of the models?

\textbf{RQ4}: Is there a discernible trade-off between model performance and CO\textsubscript2 emissions?

\subsection*{RQ1: Base Performance} \label{sec:base-performance}

In \cref{tab:main_table_results} we present the performance metrics of the reproduced models on the six datasets considered in this study. For the sake of experimental consistency, we maintain a uniform embedding size of 50 across all models. The input sequence lengths are set to 200 for FS-TKY, ML-20M, ML-1M, 100 for ML-100k and FS-NYC, and 50 for Beauty.

We verify the consistency of our results by comparing them to those reported in the original studies, although with some limitations due to differences in dataset selection between different papers. For overlapping datasets and similar experimental setups, we observe similar performance metrics, which supports the reliability of our implementation.

The first novel insight proposed by our paper is that our results do not confirm the conventional belief that SASRec consistently outperforms GRU \cite{kang2018selfattentive, sun2019bert4rec, narm, hou2022core}. Contrary to this notion, GRU4Rec beats the competitors on the 3 MovieLens datasets.

In particular, the exceptional performance of GRU4Rec in our experiments may be attributed to the quality of its implementation. As observed by \citet{hidasi2023effect}, GRU4Rec inferior performance in other studies \cite{kang2018selfattentive, sun2019bert4rec, narm, hou2022core} can be attributed to the quality of existing implementations being suboptimal. By using Torch's implementation of the GRU, our framework benefits from a high-quality implementation, which may contribute to the observed superior performance compared to other frameworks.

Furthermore, the performance of NARM and CORE is consistently below that of other models. As mentioned in \cref{sec:models}, this disparity can be attributed to the sequence-to-item training paradigm employed by these architectures, which differs from the sequence-to-sequence training employed by other models. Instead, they are able to achieve high results on datasets with short input sequences, such as Beauty, as evidenced by the best Precision and Recall @20. This finding is consistent with the observations of \citet{hou2022core}, which highlights the ability of CORE to effectively capture short-term user preferences.

\subsection*{RQ2(a): Input Sequence Length}

\begin{figure}[!t]
  \centering
  \includegraphics[width=\columnwidth]{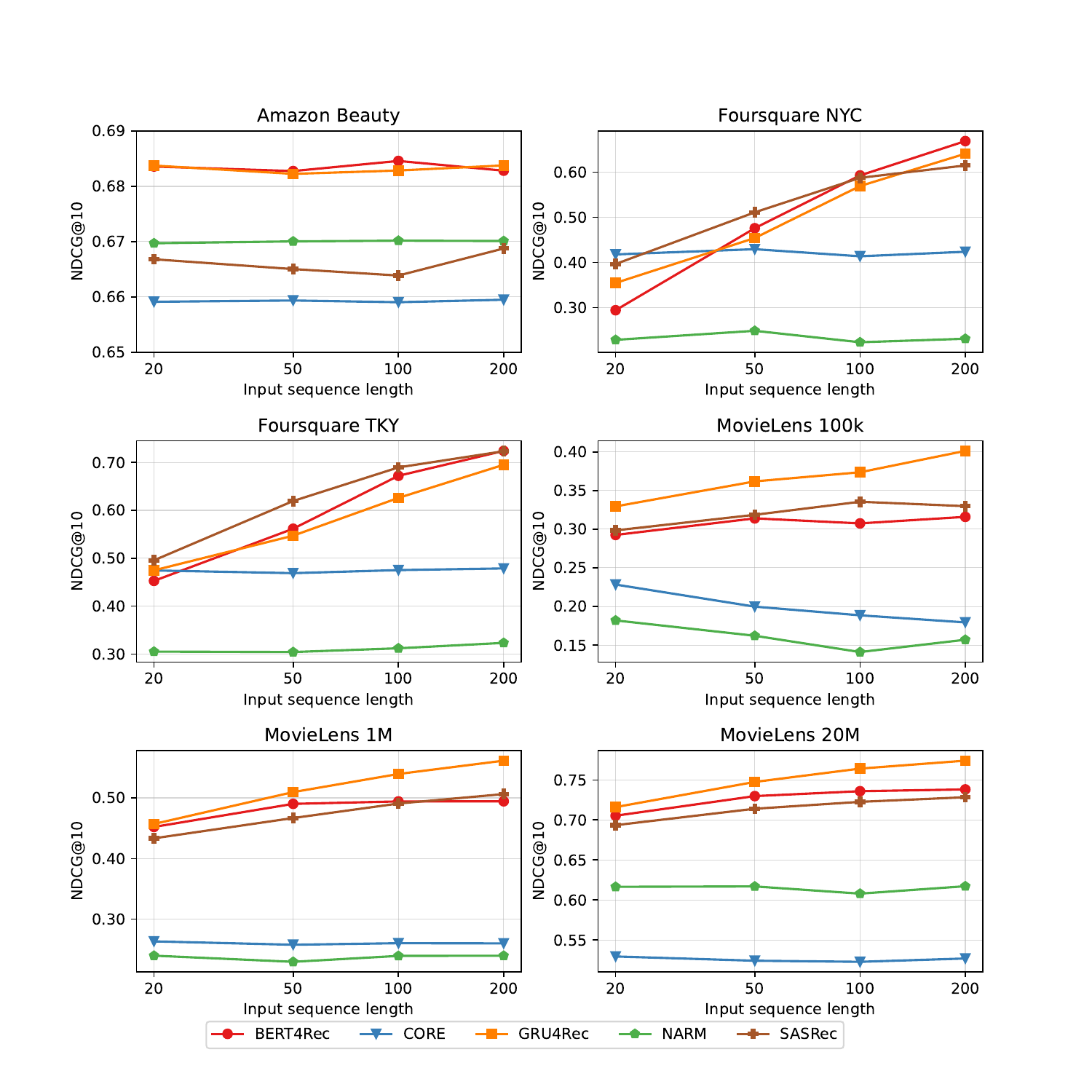}
  \caption{Effect of input sequence length on model performance, as measured by NDCG@10. Each plot shows the results of the five models on one dataset.}
  \Description[Effect of input sequence length on model performance, as measured by NDCG@10.]{Each plot shows the results of the five models on one dataset with different values of sequence length.}
  \label{fig:ablation_lookback}
\end{figure}

\cref{fig:ablation_lookback} shows the results of a detailed investigation aimed at exploring the impact of sequence length on the model's performance. In this experiment, we set the embedding size to 50 while varying the sequence length over the range of values \{20, 50, 100, 200\}.

Interestingly, for the Beauty dataset, characterized by generally short sequences, there was no significant difference in NDCG@10 when changing the input sequence length. This suggests that the models were able to maintain performance even with excessive padding. This is a positive finding, demonstrating robustness to potential noise introduced by padding in shorter sequences.

Conversely, for all other datasets, attention-based models like SASRec, BERT4Rec, and GRU4Rec, the NDCG@10 increased along with the sequence length. This finding aligns with the intuition that these models can potentially capture richer user preferences with access to a longer history of interactions.  Furthermore, it contradicts the claim made in the original SASRec paper \cite{kang2018selfattentive} that the model does not scale effectively to very long sequences. Our results suggest that SASRec may benefit from longer sequences, prompting further investigation into this aspect.

NARM and CORE exhibited contrasting behaviour. In the ML-100k dataset, their NDCG@10 scores even decreased with increasing sequence length. This is noteworthy because the original CORE paper \cite{hou2022core} only evaluated sequences up to 50 items, limiting insights into its behaviour with longer sequences. Conversely, the original NARM paper \cite{narm} claimed that NARM performs well in modeling long sessions, which aligns with our observations.

These findings highlight the importance of considering the model architecture and dataset characteristics when determining the optimal input sequence length for SRSs.

\subsection*{RQ2(b): Embedding size} \label{sec:emb_size}

A quick glance at the table could lead to the erroneous conclusion that GRU4Rec is generally the superior choice for an SRS. However, such a deduction would be deeply flawed.

\begin{figure}[!t]
  \centering
  \includegraphics[width=\columnwidth]{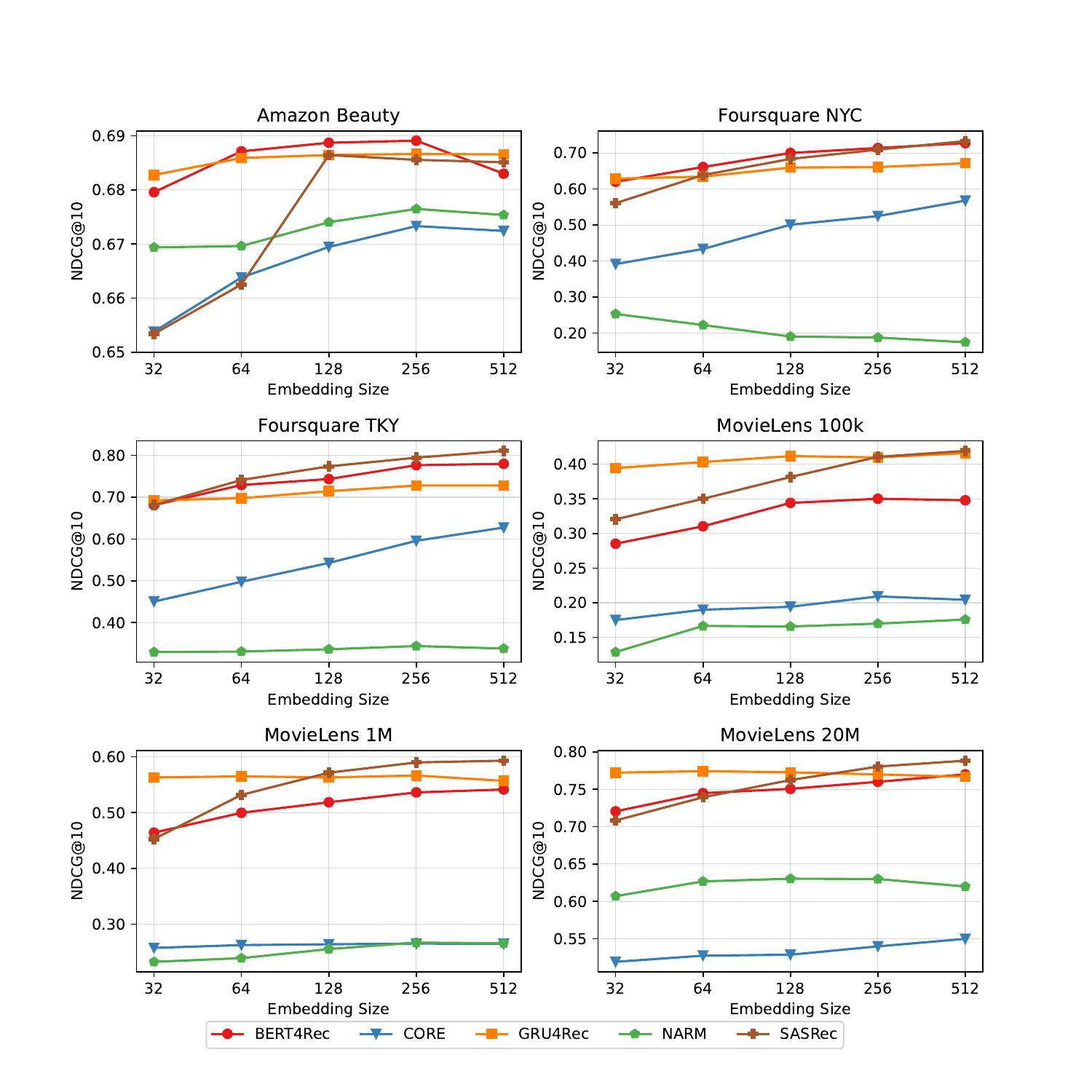}
  \caption{Effect of embedding size on model performance, as measured by NDCG@10. Each plot shows the results of the five models on one dataset.}
  \Description[Effect of embedding size on model performance, as measured by NDCG@10.]{Each plot shows the results of the five models on one dataset with different values of embedding size.}
  \label{fig:emb_ablation}
\end{figure}

\cref{fig:emb_ablation} investigates how different embedding sizes -- 32, 64, 128, 256, and 512 -- affect various models.

One striking observation is the marked influence of embedding size on SASRec and BERT4Rec. As the embedding size increases, these models show a clear improvement in performance. This phenomenon suggests that their attention mechanisms effectively exploit higher-dimensional embeddings to refine input and output representations, leading to improved predictive capabilities.

In contrast, the performance of GRU-based models, i.e. GRU4Rec and NARM, is relatively unaffected by changes in the dimensionality of the embeddings. This indicates that changes in embedding size primarily affect the number of parameters in the GRU layer without significantly impacting the computational capability.

CORE exhibits contrasting performance across different datasets. Remarkably, it demonstrates good performance on the Beauty and Foursquare datasets with increased embedding size.
However, on MovieLens datasets, CORE displays no discernible increase in performance despite variations in the embedding size.

A notable divergence from the traditional model hierarchy emerges from this analysis: transformer-based models outperform GRU-based models as embedding size increases. This suggests that, given adequate computational resources, transformer-based models offer a compelling advantage at larger embedding sizes.

Interestingly, SASRec emerges as the best-performing model when using the largest embedding size, a finding not immediately apparent from \cref{tab:main_table_results}. This highlights the nuanced relationship between model architecture, embedding size and performance.

These results highlight the importance of considering both model architecture and embedding size when designing recommendation systems, particularly in resource-rich environments where computational cost is not a limiting factor.

\subsection*{RQ3: Model Size}

\begin{figure}[!t]
    \centering
   \includegraphics[width=\columnwidth]{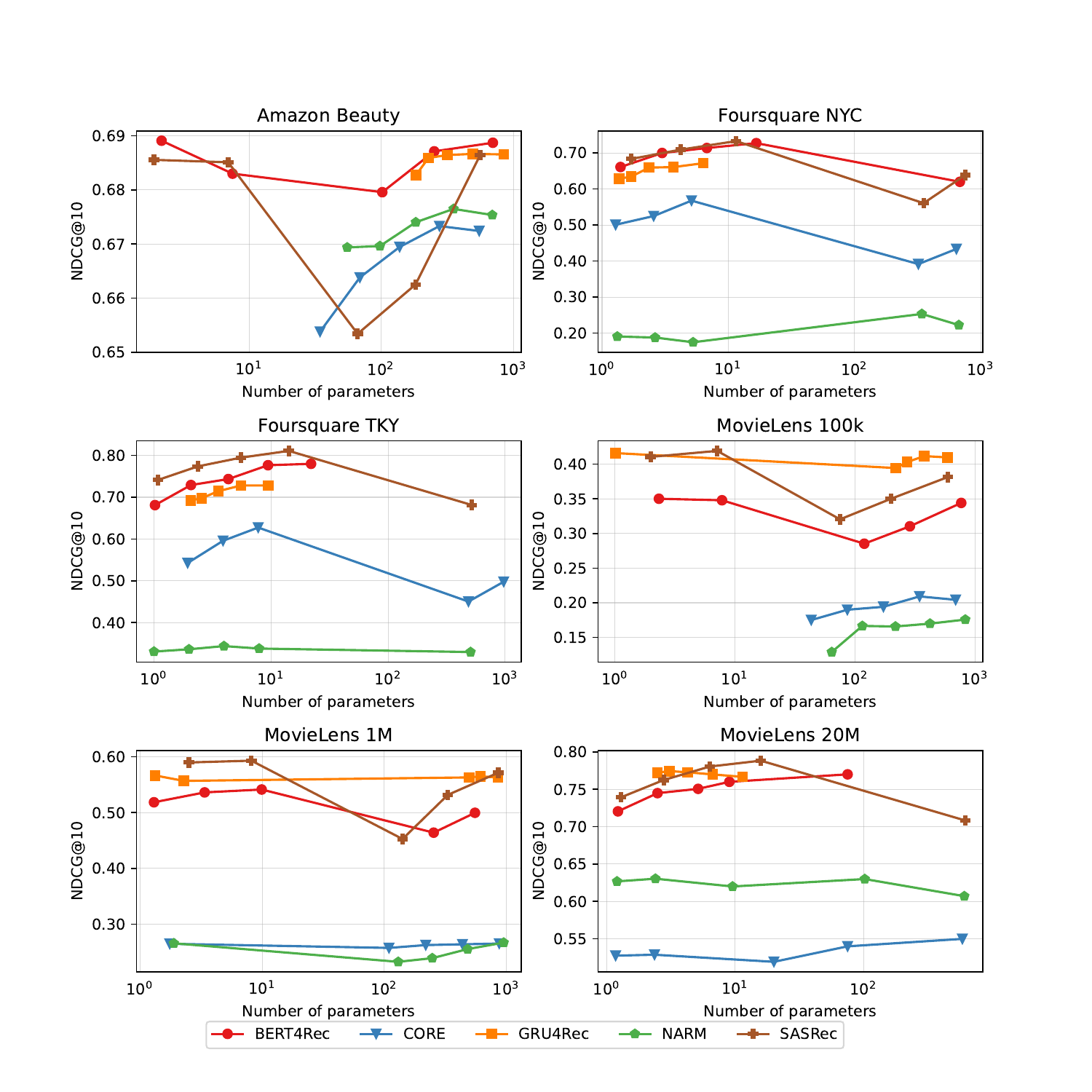}
    \caption{Effect of total number of model's parameters on the performance, as measured by NDCG@10. Each plot shows the results of the five models on one dataset.}
    \Description[Effect of number of parameters size on model performance, as measured by NDCG@10.]{Each plot shows the results of the five models on one dataset by also providing the number of parameters required by each model.}
    \label{fig:plot_number_of_params}
\end{figure}

A fundamental aspect of conducting a truly fair comparison between models is to consider their respective sizes in terms of parameters. This ensures that any differences in performance are due to inherent architectural differences and not simply to the number of parameters. \cref{fig:plot_number_of_params} shows a comparison based on this principle.

We see how the comparison becomes more nuanced when model size is considered. For example, while the number of parameters in GRU4Rec remains relatively stable even with increasing embedding sizes, transformer-based models show a contrasting trend. These models tend to grow in number of parameters with increasing embedding size more than the others.


For specific datasets, such as ML-100k and Beauty, the analysis shows that NARM still has room for improvement, as it maintains a relatively lower total number of parameters compared to other models. This suggests that NARM may have untapped potential to improve performance without significantly increasing model complexity, making it an interesting avenue for further exploration and optimisation. Regarding the Foursquare datasets, the number of parameters for NARM increases significantly when changing the embedding size, while the results in terms of NDCG remain approximately the same.

\subsection*{RQ4: Emissions}

In addition to evaluating model performance, we assess the environmental footprint of each model to identify potential trade-offs between performance and environmental impact. Our aim is to shed light on the sustainability implications associated with different model configurations.

\begin{figure}[!t]
    \centering
    \includegraphics[width=\columnwidth]{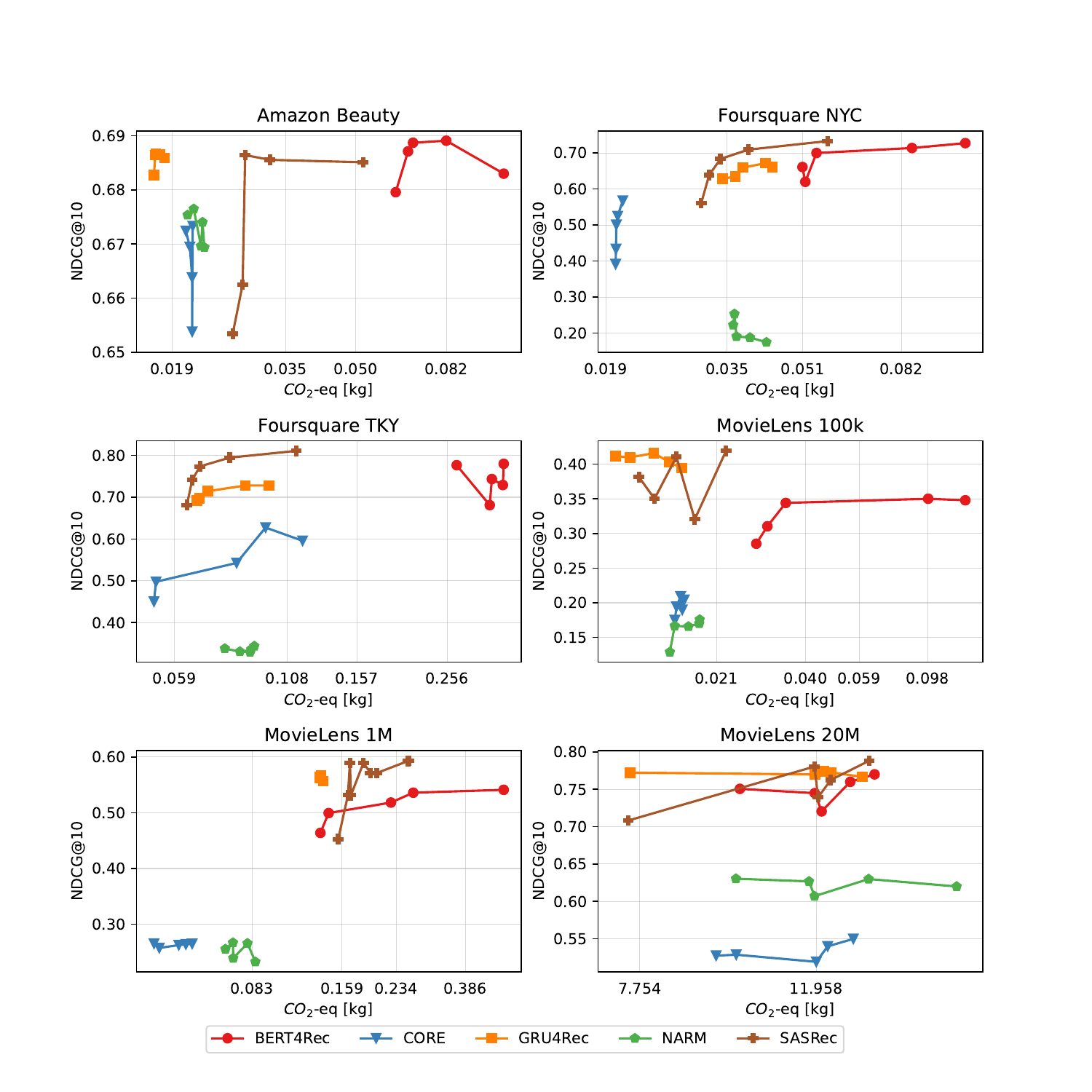}
    \caption{Relation between emissions, measured as CO\textsubscript2-eq in Kg, and performance, measured by NDCG@10. Each plot shows the results of the five models on one dataset.}
    \Description[Relation between emissions, measured as CO\textsubscript2-eq in Kg, and performance, measured by NDCG@10.]{Relation between emissions, measured as CO\textsubscript2-eq in Kg, and performance, measured by NDCG@10.}
    \label{fig:environmental_impact}
\end{figure}

\cref{fig:environmental_impact} visualises the performance as a function of the environmental impact, described by CO\textsubscript2-eq emissions in kg. A notable trend across all models is a positive correlation between CO\textsubscript2-eq emissions and performance. 
While SASRec and BERT4Rec exhibit superior performance to GRU4Rec, particularly at larger embedding sizes as discussed in \cref{sec:emb_size}, they also have a generally higher environmental impact. Transformer-based models are known for their high computational requirements, driven by their large parameter sizes \cite{10.1145/3442188.3445922}. This enhanced computational demand poses challenges in terms of hardware resources and raises concerns regarding energy consumption and environmental sustainability. In addition, it should be noted that the consumption for ML-20M is at least $10$ times higher than all other datasets.

After careful consideration, the optimal compromise appears to be the use of SASRec with an embedding size of 256. This choice strikes a balance between achieving competitive NDCG@10 values and maintaining energy consumption levels comparable to other models. This finding highlights the importance of considering environmental sustainability alongside performance metrics when selecting and optimising recommendation models.

\section{Conclusion}
Our contribution addresses reproducibility concerns in Sequential Recommendation research by introducing a standardized pre-processing and training methodology for SRSs, along with a user-friendly framework, EasyRec, for creating custom models with minimal coding effort. Our experimental results challenge existing literature by demonstrating the effectiveness of GRU4Rec with low-dimensional embeddings and the superiority of transformer-based models with larger embedding sizes. We also highlight the importance of managing input sequence length based on dataset statistics. To ensure fair model comparison, we introduce an analysis based on the number of model parameters and conduct an energy consumption analysis, identifying a trade-off between sustainability and performance. Our work represents a step towards standardizing benchmarks and pre-processing methods, promoting consistency in future research. We plan to extend our code resource with additional models, datasets, and insights to contribute to the ongoing evolution of SRS research.

\bibliographystyle{ACM-Reference-Format}
\bibliography{main}


\begin{thebibliography}{41}


\ifx \showCODEN    \undefined \def \showCODEN     #1{\unskip}     \fi
\ifx \showDOI      \undefined \def \showDOI       #1{#1}\fi
\ifx \showISBNx    \undefined \def \showISBNx     #1{\unskip}     \fi
\ifx \showISBNxiii \undefined \def \showISBNxiii  #1{\unskip}     \fi
\ifx \showISSN     \undefined \def \showISSN      #1{\unskip}     \fi
\ifx \showLCCN     \undefined \def \showLCCN      #1{\unskip}     \fi
\ifx \shownote     \undefined \def \shownote      #1{#1}          \fi
\ifx \showarticletitle \undefined \def \showarticletitle #1{#1}   \fi
\ifx \showURL      \undefined \def \showURL       {\relax}        \fi
\providecommand\bibfield[2]{#2}
\providecommand\bibinfo[2]{#2}
\providecommand\natexlab[1]{#1}
\providecommand\showeprint[2][]{arXiv:#2}

\bibitem[Anelli et~al\mbox{.}(2021)]%
        {anelli2021elliot}
\bibfield{author}{\bibinfo{person}{Vito~Walter Anelli}, \bibinfo{person}{Alejandro Bellog{\'\i}n}, \bibinfo{person}{Antonio Ferrara}, \bibinfo{person}{Daniele Malitesta}, \bibinfo{person}{Felice~Antonio Merra}, \bibinfo{person}{Claudio Pomo}, \bibinfo{person}{Francesco~Maria Donini}, {and} \bibinfo{person}{Tommaso Di~Noia}.} \bibinfo{year}{2021}\natexlab{}.
\newblock \showarticletitle{Elliot: A comprehensive and rigorous framework for reproducible recommender systems evaluation}. In \bibinfo{booktitle}{\emph{Proceedings of the 44th international ACM SIGIR conference on research and development in information retrieval}}. \bibinfo{pages}{2405--2414}.
\newblock


\bibitem[Bender et~al\mbox{.}(2021)]%
        {10.1145/3442188.3445922}
\bibfield{author}{\bibinfo{person}{Emily~M. Bender}, \bibinfo{person}{Timnit Gebru}, \bibinfo{person}{Angelina McMillan-Major}, {and} \bibinfo{person}{Shmargaret Shmitchell}.} \bibinfo{year}{2021}\natexlab{}.
\newblock \showarticletitle{On the Dangers of Stochastic Parrots: Can Language Models Be Too Big?}. In \bibinfo{booktitle}{\emph{Proceedings of the 2021 ACM Conference on Fairness, Accountability, and Transparency}} (Virtual Event, Canada) \emph{(\bibinfo{series}{FAccT '21})}. \bibinfo{publisher}{Association for Computing Machinery}, \bibinfo{address}{New York, NY, USA}, \bibinfo{pages}{610–623}.
\newblock
\showISBNx{9781450383097}
\urldef\tempurl%
\url{https://doi.org/10.1145/3442188.3445922}
\showDOI{\tempurl}


\bibitem[Beutel et~al\mbox{.}(2018)]%
        {10.1145/3159652.3159727}
\bibfield{author}{\bibinfo{person}{Alex Beutel}, \bibinfo{person}{Paul Covington}, \bibinfo{person}{Sagar Jain}, \bibinfo{person}{Can Xu}, \bibinfo{person}{Jia Li}, \bibinfo{person}{Vince Gatto}, {and} \bibinfo{person}{Ed~H. Chi}.} \bibinfo{year}{2018}\natexlab{}.
\newblock \showarticletitle{Latent Cross: Making Use of Context in Recurrent Recommender Systems}. In \bibinfo{booktitle}{\emph{Proceedings of the Eleventh ACM International Conference on Web Search and Data Mining}} (Marina Del Rey, CA, USA) \emph{(\bibinfo{series}{WSDM '18})}. \bibinfo{publisher}{Association for Computing Machinery}, \bibinfo{address}{New York, NY, USA}, \bibinfo{pages}{46–54}.
\newblock
\showISBNx{9781450355810}
\urldef\tempurl%
\url{https://doi.org/10.1145/3159652.3159727}
\showDOI{\tempurl}


\bibitem[Cho et~al\mbox{.}(2014)]%
        {cho2014learning}
\bibfield{author}{\bibinfo{person}{Kyunghyun Cho}, \bibinfo{person}{Bart van Merrienboer}, \bibinfo{person}{Caglar Gulcehre}, \bibinfo{person}{Dzmitry Bahdanau}, \bibinfo{person}{Fethi Bougares}, \bibinfo{person}{Holger Schwenk}, {and} \bibinfo{person}{Yoshua Bengio}.} \bibinfo{year}{2014}\natexlab{}.
\newblock \bibinfo{title}{Learning Phrase Representations using RNN Encoder-Decoder for Statistical Machine Translation}.
\newblock
\newblock
\showeprint[arxiv]{1406.1078}~[cs.CL]


\bibitem[Courty et~al\mbox{.}(2023)]%
        {codecarbon}
\bibfield{author}{\bibinfo{person}{Benoit Courty}, \bibinfo{person}{Victor Schmidt}, \bibinfo{person}{Goyal-Kamal}, \bibinfo{person}{MarionCoutarel}, \bibinfo{person}{Boris Feld}, \bibinfo{person}{Jérémy Lecourt}, \bibinfo{person}{SabAmine}, \bibinfo{person}{kngoyal}, \bibinfo{person}{Mathilde Léval}, \bibinfo{person}{Alexis Cruveiller}, \bibinfo{person}{inimaz}, \bibinfo{person}{ouminasara}, \bibinfo{person}{Franklin Zhao}, \bibinfo{person}{Aditya Joshi}, \bibinfo{person}{Alexis Bogroff}, \bibinfo{person}{Amine Saboni}, \bibinfo{person}{Hugues de Lavoreille}, \bibinfo{person}{Niko Laskaris}, \bibinfo{person}{Luis Blanche}, \bibinfo{person}{Edoardo Abati}, \bibinfo{person}{LiamConnell}, \bibinfo{person}{Douglas Blank}, \bibinfo{person}{Ziyao Wang}, \bibinfo{person}{Armin Catovic}, \bibinfo{person}{Michał Stęchły}, \bibinfo{person}{alencon}, \bibinfo{person}{JPW}, \bibinfo{person}{MinervaBooks}, \bibinfo{person}{Necmettin Çarkacı}, {and} \bibinfo{person}{DomAlexRod}.} \bibinfo{year}{2023}\natexlab{}.
\newblock \bibinfo{title}{mlco2/codecarbon: v2.3.2}.
\newblock
\newblock
\urldef\tempurl%
\url{https://doi.org/10.5281/zenodo.10213072}
\showDOI{\tempurl}


\bibitem[Cremonesi and Jannach(2021)]%
        {cremonesi2021progress}
\bibfield{author}{\bibinfo{person}{Paolo Cremonesi} {and} \bibinfo{person}{Dietmar Jannach}.} \bibinfo{year}{2021}\natexlab{}.
\newblock \showarticletitle{Progress in recommender systems research: Crisis? What crisis?}
\newblock \bibinfo{journal}{\emph{AI Magazine}} \bibinfo{volume}{42}, \bibinfo{number}{3} (\bibinfo{year}{2021}), \bibinfo{pages}{43--54}.
\newblock


\bibitem[Devlin et~al\mbox{.}(2019)]%
        {devlin-etal-2019-bert}
\bibfield{author}{\bibinfo{person}{Jacob Devlin}, \bibinfo{person}{Ming-Wei Chang}, \bibinfo{person}{Kenton Lee}, {and} \bibinfo{person}{Kristina Toutanova}.} \bibinfo{year}{2019}\natexlab{}.
\newblock \showarticletitle{{BERT}: Pre-training of Deep Bidirectional Transformers for Language Understanding}. In \bibinfo{booktitle}{\emph{Proceedings of the 2019 Conference of the North {A}merican Chapter of the Association for Computational Linguistics: Human Language Technologies, Volume 1 (Long and Short Papers)}}, \bibfield{editor}{\bibinfo{person}{Jill Burstein}, \bibinfo{person}{Christy Doran}, {and} \bibinfo{person}{Thamar Solorio}} (Eds.). \bibinfo{publisher}{Association for Computational Linguistics}, \bibinfo{address}{Minneapolis, Minnesota}, \bibinfo{pages}{4171--4186}.
\newblock
\urldef\tempurl%
\url{https://doi.org/10.18653/v1/N19-1423}
\showDOI{\tempurl}


\bibitem[Ferrari~Dacrema et~al\mbox{.}(2021)]%
        {ferrari2021troubling}
\bibfield{author}{\bibinfo{person}{Maurizio Ferrari~Dacrema}, \bibinfo{person}{Simone Boglio}, \bibinfo{person}{Paolo Cremonesi}, {and} \bibinfo{person}{Dietmar Jannach}.} \bibinfo{year}{2021}\natexlab{}.
\newblock \showarticletitle{A troubling analysis of reproducibility and progress in recommender systems research}.
\newblock \bibinfo{journal}{\emph{ACM Transactions on Information Systems (TOIS)}} \bibinfo{volume}{39}, \bibinfo{number}{2} (\bibinfo{year}{2021}), \bibinfo{pages}{1--49}.
\newblock


\bibitem[Ferrari~Dacrema et~al\mbox{.}(2019)]%
        {ferrari2019we}
\bibfield{author}{\bibinfo{person}{Maurizio Ferrari~Dacrema}, \bibinfo{person}{Paolo Cremonesi}, {and} \bibinfo{person}{Dietmar Jannach}.} \bibinfo{year}{2019}\natexlab{}.
\newblock \bibinfo{title}{Are we really making much progress? A worrying analysis of recent neural recommendation approaches}.
\newblock , \bibinfo{numpages}{101--109}~pages.
\newblock


\bibitem[Ferrari~Dacrema et~al\mbox{.}(2020)]%
        {ferrari2020methodological}
\bibfield{author}{\bibinfo{person}{Maurizio Ferrari~Dacrema}, \bibinfo{person}{Paolo Cremonesi}, \bibinfo{person}{Dietmar Jannach}, {et~al\mbox{.}}} \bibinfo{year}{2020}\natexlab{}.
\newblock \bibinfo{title}{Methodological issues in recommender systems research}.
\newblock , \bibinfo{numpages}{4706--4710}~pages.
\newblock


\bibitem[Good(1952)]%
        {good1952rational}
\bibfield{author}{\bibinfo{person}{Irving~John Good}.} \bibinfo{year}{1952}\natexlab{}.
\newblock \showarticletitle{Rational decisions}.
\newblock \bibinfo{journal}{\emph{Journal of the Royal Statistical Society: Series B (Methodological)}} \bibinfo{volume}{14}, \bibinfo{number}{1} (\bibinfo{year}{1952}), \bibinfo{pages}{107--114}.
\newblock


\bibitem[Harper and Konstan(2015)]%
        {10.1145/2827872}
\bibfield{author}{\bibinfo{person}{F.~Maxwell Harper} {and} \bibinfo{person}{Joseph~A. Konstan}.} \bibinfo{year}{2015}\natexlab{}.
\newblock \showarticletitle{The MovieLens Datasets: History and Context}.
\newblock \bibinfo{journal}{\emph{ACM Trans. Interact. Intell. Syst.}} \bibinfo{volume}{5}, \bibinfo{number}{4}, Article \bibinfo{articleno}{19} (\bibinfo{date}{dec} \bibinfo{year}{2015}), \bibinfo{numpages}{19}~pages.
\newblock
\showISSN{2160-6455}
\urldef\tempurl%
\url{https://doi.org/10.1145/2827872}
\showDOI{\tempurl}


\bibitem[Hidasi and Czapp(2023a)]%
        {hidasi2023effect}
\bibfield{author}{\bibinfo{person}{Bal{\'a}zs Hidasi} {and} \bibinfo{person}{{\'A}d{\'a}m~Tibor Czapp}.} \bibinfo{year}{2023}\natexlab{a}.
\newblock \bibinfo{title}{The effect of third party implementations on reproducibility}.
\newblock , \bibinfo{numpages}{272--282}~pages.
\newblock


\bibitem[Hidasi and Czapp(2023b)]%
        {hidasi2023widespread}
\bibfield{author}{\bibinfo{person}{Bal{\'a}zs Hidasi} {and} \bibinfo{person}{{\'A}d{\'a}m~Tibor Czapp}.} \bibinfo{year}{2023}\natexlab{b}.
\newblock \bibinfo{title}{Widespread Flaws in Offline Evaluation of Recommender Systems}.
\newblock , \bibinfo{numpages}{848--855}~pages.
\newblock


\bibitem[Hidasi et~al\mbox{.}(2016)]%
        {hidasi2016sessionbased}
\bibfield{author}{\bibinfo{person}{Balázs Hidasi}, \bibinfo{person}{Alexandros Karatzoglou}, \bibinfo{person}{Linas Baltrunas}, {and} \bibinfo{person}{Domonkos Tikk}.} \bibinfo{year}{2016}\natexlab{}.
\newblock \bibinfo{title}{Session-based Recommendations with Recurrent Neural Networks}.
\newblock
\newblock
\showeprint[arxiv]{1511.06939}~[cs.LG]


\bibitem[Hou et~al\mbox{.}(2022)]%
        {hou2022core}
\bibfield{author}{\bibinfo{person}{Yupeng Hou}, \bibinfo{person}{Binbin Hu}, \bibinfo{person}{Zhiqiang Zhang}, {and} \bibinfo{person}{Wayne~Xin Zhao}.} \bibinfo{year}{2022}\natexlab{}.
\newblock \bibinfo{title}{Core: simple and effective session-based recommendation within consistent representation space}.
\newblock , \bibinfo{numpages}{1796--1801}~pages.
\newblock


\bibitem[Hunter et~al\mbox{.}(2021)]%
        {hunter2021glasgow}
\bibfield{author}{\bibinfo{person}{David~B Hunter}, \bibinfo{person}{James~E Salzman}, {and} \bibinfo{person}{Durwood Zaelke}.} \bibinfo{year}{2021}\natexlab{}.
\newblock \bibinfo{title}{Glasgow climate summit: Cop26}.
\newblock
\newblock


\bibitem[Kang and McAuley(2018)]%
        {kang2018selfattentive}
\bibfield{author}{\bibinfo{person}{Wang-Cheng Kang} {and} \bibinfo{person}{Julian McAuley}.} \bibinfo{year}{2018}\natexlab{}.
\newblock \bibinfo{title}{Self-Attentive Sequential Recommendation}.
\newblock
\newblock
\showeprint[arxiv]{1808.09781}~[cs.IR]


\bibitem[Kim and Neff(2009)]%
        {kim2009measurement}
\bibfield{author}{\bibinfo{person}{Brent Kim} {and} \bibinfo{person}{Roni Neff}.} \bibinfo{year}{2009}\natexlab{}.
\newblock \showarticletitle{Measurement and communication of greenhouse gas emissions from US food consumption via carbon calculators}.
\newblock \bibinfo{journal}{\emph{Ecological Economics}} \bibinfo{volume}{69}, \bibinfo{number}{1} (\bibinfo{year}{2009}), \bibinfo{pages}{186--196}.
\newblock


\bibitem[Klenitskiy and Vasilev(2023)]%
        {klenitskiy2023turning}
\bibfield{author}{\bibinfo{person}{Anton Klenitskiy} {and} \bibinfo{person}{Alexey Vasilev}.} \bibinfo{year}{2023}\natexlab{}.
\newblock \bibinfo{title}{Turning Dross Into Gold Loss: is BERT4Rec really better than SASRec?}
\newblock , \bibinfo{numpages}{1120--1125}~pages.
\newblock


\bibitem[Li et~al\mbox{.}(2017)]%
        {narm}
\bibfield{author}{\bibinfo{person}{Jing Li}, \bibinfo{person}{Pengjie Ren}, \bibinfo{person}{Zhumin Chen}, \bibinfo{person}{Zhaochun Ren}, \bibinfo{person}{Tao Lian}, {and} \bibinfo{person}{Jun Ma}.} \bibinfo{year}{2017}\natexlab{}.
\newblock \showarticletitle{Neural Attentive Session-based Recommendation}. In \bibinfo{booktitle}{\emph{Proceedings of the 2017 ACM on Conference on Information and Knowledge Management}} (Singapore, Singapore) \emph{(\bibinfo{series}{CIKM '17})}. \bibinfo{publisher}{Association for Computing Machinery}, \bibinfo{address}{New York, NY, USA}, \bibinfo{pages}{1419–1428}.
\newblock
\showISBNx{9781450349185}
\urldef\tempurl%
\url{https://doi.org/10.1145/3132847.3132926}
\showDOI{\tempurl}


\bibitem[Li et~al\mbox{.}(2023)]%
        {li2023repetition}
\bibfield{author}{\bibinfo{person}{Ming Li}, \bibinfo{person}{Ali Vardasbi}, \bibinfo{person}{Andrew Yates}, {and} \bibinfo{person}{Maarten de Rijke}.} \bibinfo{year}{2023}\natexlab{}.
\newblock \bibinfo{title}{Repetition and Exploration in Sequential Recommendation}.
\newblock
\newblock


\bibitem[Lin(2019)]%
        {lin2019neural}
\bibfield{author}{\bibinfo{person}{Jimmy Lin}.} \bibinfo{year}{2019}\natexlab{}.
\newblock \bibinfo{title}{The neural hype and comparisons against weak baselines}.
\newblock , \bibinfo{numpages}{40--51}~pages.
\newblock


\bibitem[Ludewig and Jannach(2018)]%
        {ludewig2018evaluation}
\bibfield{author}{\bibinfo{person}{Malte Ludewig} {and} \bibinfo{person}{Dietmar Jannach}.} \bibinfo{year}{2018}\natexlab{}.
\newblock \showarticletitle{Evaluation of session-based recommendation algorithms}.
\newblock \bibinfo{journal}{\emph{User Modeling and User-Adapted Interaction}}  \bibinfo{volume}{28} (\bibinfo{year}{2018}), \bibinfo{pages}{331--390}.
\newblock


\bibitem[McAuley et~al\mbox{.}(2015)]%
        {10.1145/2766462.2767755}
\bibfield{author}{\bibinfo{person}{Julian McAuley}, \bibinfo{person}{Christopher Targett}, \bibinfo{person}{Qinfeng Shi}, {and} \bibinfo{person}{Anton van~den Hengel}.} \bibinfo{year}{2015}\natexlab{}.
\newblock \showarticletitle{Image-Based Recommendations on Styles and Substitutes}. In \bibinfo{booktitle}{\emph{Proceedings of the 38th International ACM SIGIR Conference on Research and Development in Information Retrieval}} (Santiago, Chile) \emph{(\bibinfo{series}{SIGIR '15})}. \bibinfo{publisher}{Association for Computing Machinery}, \bibinfo{address}{New York, NY, USA}, \bibinfo{pages}{43–52}.
\newblock
\showISBNx{9781450336215}
\urldef\tempurl%
\url{https://doi.org/10.1145/2766462.2767755}
\showDOI{\tempurl}


\bibitem[Michiels et~al\mbox{.}(2022)]%
        {recpack2022}
\bibfield{author}{\bibinfo{person}{Lien Michiels}, \bibinfo{person}{Robin Verachtert}, {and} \bibinfo{person}{Bart Goethals}.} \bibinfo{year}{2022}\natexlab{}.
\newblock \showarticletitle{RecPack: An(Other) Experimentation Toolkit for Top-N Recommendation Using Implicit Feedback Data}. In \bibinfo{booktitle}{\emph{Proceedings of the 16th ACM Conference on Recommender Systems}} (Seattle, WA, USA) \emph{(\bibinfo{series}{RecSys '22})}. \bibinfo{publisher}{Association for Computing Machinery}, \bibinfo{address}{New York, NY, USA}, \bibinfo{pages}{648–651}.
\newblock
\showISBNx{9781450392785}
\urldef\tempurl%
\url{https://doi.org/10.1145/3523227.3551472}
\showDOI{\tempurl}


\bibitem[Pascanu et~al\mbox{.}(2013)]%
        {pascanu2013difficulty}
\bibfield{author}{\bibinfo{person}{Razvan Pascanu}, \bibinfo{person}{Tomas Mikolov}, {and} \bibinfo{person}{Yoshua Bengio}.} \bibinfo{year}{2013}\natexlab{}.
\newblock \bibinfo{title}{On the difficulty of training Recurrent Neural Networks}.
\newblock
\newblock
\showeprint[arxiv]{1211.5063}~[cs.LG]


\bibitem[Patterson et~al\mbox{.}(2021)]%
        {patterson2021carbon}
\bibfield{author}{\bibinfo{person}{David Patterson}, \bibinfo{person}{Joseph Gonzalez}, \bibinfo{person}{Quoc Le}, \bibinfo{person}{Chen Liang}, \bibinfo{person}{Lluis-Miquel Munguia}, \bibinfo{person}{Daniel Rothchild}, \bibinfo{person}{David So}, \bibinfo{person}{Maud Texier}, {and} \bibinfo{person}{Jeff Dean}.} \bibinfo{year}{2021}\natexlab{}.
\newblock \bibinfo{title}{Carbon emissions and large neural network training}.
\newblock
\newblock


\bibitem[Petrov and Macdonald(2022)]%
        {petrov2022systematic}
\bibfield{author}{\bibinfo{person}{Aleksandr Petrov} {and} \bibinfo{person}{Craig Macdonald}.} \bibinfo{year}{2022}\natexlab{}.
\newblock \bibinfo{title}{A systematic review and replicability study of bert4rec for sequential recommendation}.
\newblock , \bibinfo{numpages}{436--447}~pages.
\newblock


\bibitem[Quadrana et~al\mbox{.}(2018)]%
        {quadrana2018sequence}
\bibfield{author}{\bibinfo{person}{Massimo Quadrana}, \bibinfo{person}{Paolo Cremonesi}, {and} \bibinfo{person}{Dietmar Jannach}.} \bibinfo{year}{2018}\natexlab{}.
\newblock \showarticletitle{Sequence-aware recommender systems}.
\newblock \bibinfo{journal}{\emph{ACM Computing Surveys (CSUR)}} \bibinfo{volume}{51}, \bibinfo{number}{4} (\bibinfo{year}{2018}), \bibinfo{pages}{1--36}.
\newblock


\bibitem[Rendle et~al\mbox{.}(2010)]%
        {10.1145/1772690.1772773}
\bibfield{author}{\bibinfo{person}{Steffen Rendle}, \bibinfo{person}{Christoph Freudenthaler}, {and} \bibinfo{person}{Lars Schmidt-Thieme}.} \bibinfo{year}{2010}\natexlab{}.
\newblock \showarticletitle{Factorizing personalized Markov chains for next-basket recommendation}. In \bibinfo{booktitle}{\emph{Proceedings of the 19th International Conference on World Wide Web}} (Raleigh, North Carolina, USA) \emph{(\bibinfo{series}{WWW '10})}. \bibinfo{publisher}{Association for Computing Machinery}, \bibinfo{address}{New York, NY, USA}, \bibinfo{pages}{811–820}.
\newblock
\showISBNx{9781605587998}
\urldef\tempurl%
\url{https://doi.org/10.1145/1772690.1772773}
\showDOI{\tempurl}


\bibitem[Salah et~al\mbox{.}(2020)]%
        {salah2020cornac}
\bibfield{author}{\bibinfo{person}{Aghiles Salah}, \bibinfo{person}{Quoc-Tuan Truong}, {and} \bibinfo{person}{Hady~W Lauw}.} \bibinfo{year}{2020}\natexlab{}.
\newblock \showarticletitle{Cornac: A Comparative Framework for Multimodal Recommender Systems}.
\newblock \bibinfo{journal}{\emph{Journal of Machine Learning Research}} \bibinfo{volume}{21}, \bibinfo{number}{95} (\bibinfo{year}{2020}), \bibinfo{pages}{1--5}.
\newblock


\bibitem[Smith and Linden(2017)]%
        {7927889}
\bibfield{author}{\bibinfo{person}{Brent Smith} {and} \bibinfo{person}{Greg Linden}.} \bibinfo{year}{2017}\natexlab{}.
\newblock \showarticletitle{Two Decades of Recommender Systems at Amazon.com}.
\newblock \bibinfo{journal}{\emph{IEEE Internet Computing}} \bibinfo{volume}{21}, \bibinfo{number}{3} (\bibinfo{year}{2017}), \bibinfo{pages}{12--18}.
\newblock
\urldef\tempurl%
\url{https://doi.org/10.1109/MIC.2017.72}
\showDOI{\tempurl}


\bibitem[Sun et~al\mbox{.}(2019)]%
        {sun2019bert4rec}
\bibfield{author}{\bibinfo{person}{Fei Sun}, \bibinfo{person}{Jun Liu}, \bibinfo{person}{Jian Wu}, \bibinfo{person}{Changhua Pei}, \bibinfo{person}{Xiao Lin}, \bibinfo{person}{Wenwu Ou}, {and} \bibinfo{person}{Peng Jiang}.} \bibinfo{year}{2019}\natexlab{}.
\newblock \bibinfo{title}{BERT4Rec: Sequential Recommendation with Bidirectional Encoder Representations from Transformer}.
\newblock
\newblock
\showeprint[arxiv]{1904.06690}~[cs.IR]


\bibitem[Sun et~al\mbox{.}(2020)]%
        {sun2020we}
\bibfield{author}{\bibinfo{person}{Zhu Sun}, \bibinfo{person}{Di Yu}, \bibinfo{person}{Hui Fang}, \bibinfo{person}{Jie Yang}, \bibinfo{person}{Xinghua Qu}, \bibinfo{person}{Jie Zhang}, {and} \bibinfo{person}{Cong Geng}.} \bibinfo{year}{2020}\natexlab{}.
\newblock \bibinfo{title}{Are we evaluating rigorously? benchmarking recommendation for reproducible evaluation and fair comparison}.
\newblock , \bibinfo{numpages}{23--32}~pages.
\newblock


\bibitem[Tang and Wang(2018)]%
        {10.1145/3159652.3159656}
\bibfield{author}{\bibinfo{person}{Jiaxi Tang} {and} \bibinfo{person}{Ke Wang}.} \bibinfo{year}{2018}\natexlab{}.
\newblock \showarticletitle{Personalized Top-N Sequential Recommendation via Convolutional Sequence Embedding}. In \bibinfo{booktitle}{\emph{Proceedings of the Eleventh ACM International Conference on Web Search and Data Mining}} (Marina Del Rey, CA, USA) \emph{(\bibinfo{series}{WSDM '18})}. \bibinfo{publisher}{Association for Computing Machinery}, \bibinfo{address}{New York, NY, USA}, \bibinfo{pages}{565–573}.
\newblock
\showISBNx{9781450355810}
\urldef\tempurl%
\url{https://doi.org/10.1145/3159652.3159656}
\showDOI{\tempurl}


\bibitem[Vaswani et~al\mbox{.}(2017)]%
        {vaswani2017attention}
\bibfield{author}{\bibinfo{person}{Ashish Vaswani}, \bibinfo{person}{Noam Shazeer}, \bibinfo{person}{Niki Parmar}, \bibinfo{person}{Jakob Uszkoreit}, \bibinfo{person}{Llion Jones}, \bibinfo{person}{Aidan~N Gomez}, \bibinfo{person}{{\L}ukasz Kaiser}, {and} \bibinfo{person}{Illia Polosukhin}.} \bibinfo{year}{2017}\natexlab{}.
\newblock \showarticletitle{Attention is all you need}.
\newblock \bibinfo{journal}{\emph{Advances in neural information processing systems}}  \bibinfo{volume}{30} (\bibinfo{year}{2017}).
\newblock


\bibitem[Wang et~al\mbox{.}(2020)]%
        {wang2020rechorus}
\bibfield{author}{\bibinfo{person}{Chenyang Wang}, \bibinfo{person}{Min Zhang}, \bibinfo{person}{Weizhi Ma}, \bibinfo{person}{Yiqun Liu}, {and} \bibinfo{person}{Shaoping Ma}.} \bibinfo{year}{2020}\natexlab{}.
\newblock \showarticletitle{Make it a chorus: knowledge-and time-aware item modeling for sequential recommendation}. In \bibinfo{booktitle}{\emph{Proceedings of the 43rd International ACM SIGIR conference on research and development in Information Retrieval}}. \bibinfo{pages}{109--118}.
\newblock


\bibitem[Wang et~al\mbox{.}(2019)]%
        {Wang_2019}
\bibfield{author}{\bibinfo{person}{Shoujin Wang}, \bibinfo{person}{Liang Hu}, \bibinfo{person}{Yan Wang}, \bibinfo{person}{Longbing Cao}, \bibinfo{person}{Quan~Z. Sheng}, {and} \bibinfo{person}{Mehmet Orgun}.} \bibinfo{year}{2019}\natexlab{}.
\newblock \bibinfo{title}{Sequential Recommender Systems: Challenges, Progress and Prospects}.
\newblock
\newblock
\urldef\tempurl%
\url{https://doi.org/10.24963/ijcai.2019/883}
\showDOI{\tempurl}


\bibitem[Yang et~al\mbox{.}(2015)]%
        {6844862}
\bibfield{author}{\bibinfo{person}{Dingqi Yang}, \bibinfo{person}{Daqing Zhang}, \bibinfo{person}{Vincent~W. Zheng}, {and} \bibinfo{person}{Zhiyong Yu}.} \bibinfo{year}{2015}\natexlab{}.
\newblock \showarticletitle{Modeling User Activity Preference by Leveraging User Spatial Temporal Characteristics in LBSNs}.
\newblock \bibinfo{journal}{\emph{IEEE Transactions on Systems, Man, and Cybernetics: Systems}} \bibinfo{volume}{45}, \bibinfo{number}{1} (\bibinfo{year}{2015}), \bibinfo{pages}{129--142}.
\newblock
\urldef\tempurl%
\url{https://doi.org/10.1109/TSMC.2014.2327053}
\showDOI{\tempurl}


\bibitem[Zhao et~al\mbox{.}(2021)]%
        {zhao2021recbole}
\bibfield{author}{\bibinfo{person}{Wayne~Xin Zhao}, \bibinfo{person}{Shanlei Mu}, \bibinfo{person}{Yupeng Hou}, \bibinfo{person}{Zihan Lin}, \bibinfo{person}{Yushuo Chen}, \bibinfo{person}{Xingyu Pan}, \bibinfo{person}{Kaiyuan Li}, \bibinfo{person}{Yujie Lu}, \bibinfo{person}{Hui Wang}, \bibinfo{person}{Changxin Tian}, \bibinfo{person}{Yingqian Min}, \bibinfo{person}{Zhichao Feng}, \bibinfo{person}{Xinyan Fan}, \bibinfo{person}{Xu Chen}, \bibinfo{person}{Pengfei Wang}, \bibinfo{person}{Wendi Ji}, \bibinfo{person}{Yaliang Li}, \bibinfo{person}{Xiaoling Wang}, {and} \bibinfo{person}{Ji-Rong Wen}.} \bibinfo{year}{2021}\natexlab{}.
\newblock \bibinfo{title}{RecBole: Towards a Unified, Comprehensive and Efficient Framework for Recommendation Algorithms}.
\newblock
\newblock
\showeprint[arxiv]{2011.01731}~[cs.IR]


\end{thebibliography}
\end{document}